\documentclass[final] {aipproc}
\usepackage{sidecap,graphicx,boxedminipage,epsfig,wrapfig,floatflt,shadow}

\layoutstyle{8x11double}

\begin{document}

\title{Transfer of Learning in Quantum Mechanics}


\author{Chandralekha Singh}{
  address={Department of Physics, University of Pittsburgh, Pittsburgh, PA, 15260}
}

\begin{abstract}
We investigate the difficulties that undergraduate students in quantum mechanics courses have in transferring learning from
previous courses or within the same course from one context to another by administering written tests and conducting individual interviews.
Quantum mechanics is abstract and its paradigm is very different from the classical one. A good grasp of the principles of quantum mechanics 
requires creating and organizing a knowledge structure consistent with the quantum postulates.
Previously learned concepts such as the principle of superposition and probability can be useful in quantum mechanics if students are given 
opportunity to build associations between new and prior knowledge.
We also discuss the need for better alignment between quantum mechanics and modern physics courses taken previously
because semi-classical models can impede internalization of the quantum paradigm in more advanced courses.
\end{abstract}

\maketitle

\section{Introduction}

Transfer of learning is defined as the application of knowledge and skill acquired in one context to another~\cite{gick}. Cognitive theory suggests that transfer
can be difficult especially if the source (from which knowledge is to be transferred) and target (to which knowledge is to be transferred) do not 
share surface features. The failure to transfer is attributed to the fact that knowledge is encoded in memory with the context in which it was acquired
and the features of the target may not trigger the recall of relevant resources in memory about the source even if the two share deep features.
Transfer in physics is often challenging. There are very few concepts and principles in physics and learning requires unpacking and understanding
their applicability in diverse physical situations. For example, waves can be mechanical, electromagnetic or quantum mechanical, all of which share deep
features but are superficially very different. However, transfer is essential for a good grasp of physics because physics is hierarchical and new
knowledge builds on prior knowledge. In order to improve the transfer, it is important to help students de-contextualize learning and store it in 
memory at an abstract level. This can be facilitated by learning the same concept in different contexts.

It is clear that some appropriate transfer does occur from other courses to quantum mechanics (QM). Here, we will focus on identifying learning that is
difficult to transfer appropriately for students in QM and discuss possible reasons.  This type of study is important for designing
effective instructional strategies to improve transfer~\cite{singh}. Our investigation involves administering and analyzing
written tests and conducting individual interviews.
In this paper, we will outline the general difficulties found and their possible sources. Detailed statistics will be presented elsewhere.

Sources of major difficulties with QM include its novel paradigm, abstractness and mathematical sophistication. The quantum paradigm 
does not conform to our everyday experience.
In order to transfer previous learning, e.g., waves or probability concepts learned in classical contexts, students must first learn the 
basic structure of quantum mechanics and then contemplate how the previously learned knowledge applies to this novel framework.
Also, most students take a one or two semester course in modern physics before QM in which quantum mechanics is covered. Below, we discuss why
there should be better co-ordination between the teaching of modern physics and quantum mechanics, and
more effort channeled on helping students understand the limits of the semi-classical models and perhaps the 
time alloted to such models should be reduced.

\section{Inappropriate Transfer from Classical Mechanics}

The written free-response questions and individual interviews show that some students have not acquired or retained relevant knowledge 
from previous courses (e.g., superposition principle).
Here, we will focus on evidence pointing to the difficulty in appropriate transfer
due to lack of robust knowledge hierarchy (schema) for quantum mechanics. For example, one common difficulty is assuming that an object with a label 
``x" is orthogonal to or cannot influence an object with a label ``y". This is evident from responses such as ``The magnetic field is in the 
z direction so electron is not influenced if it is initially in an eigenstate of $S_x$" or ``Eigenstates of $S_x$ are orthogonal to eigenstates 
of $S_y$". In introductory physics, x, y and z are indeed conventional labels for orthogonal components of a vector. Unless we give students
an opportunity to understand the structure of quantum mechanics, such difficulties will persist. Students must be given opportunity to build schemas
that allows them to understand that although an electron in an external magnetic field pointing
in the z direction is in a real physical three dimensional space of the lab, to make predictions about the measurement performed in the lab
using quantum mechanics, one needs to map the problem to an abstract 
Hilbert space in which the state of the system lies and
where all the observables of the real physical space get mapped on to operators acting on states. 

The conflict between the classical and quantum paradigm also forced students to incorrectly conclude that ``Successive measurements of continuous
variables of a particle, e.g., position, produce ``somewhat" deterministic outcomes whereas successive measurements of discrete variables, e.g., spin,
can produce very different outcomes." For example, during the interview, one students said the following regarding successive measurements on 
a single electron: ``If an observable has a continuous 
spectrum...the next measurement won't be very different from the first one. But if the spectrum is discrete then you will get very 
different outcomes." When asked to
elaborate the student added ``For example, imagine measuring the position of an electron. It is a continuous function so that time
dependence is gentle and after a few seconds you can only go from A to its neighboring point (pointing to an x vs. t graph that he
sketches on the paper) you cannot go from this without going through this intermediate space." When asked to elaborate his views
about observables with discrete spectrum, the student said ``Think of discrete variables like spin...they can give you very different
values in a short time because the system must flip from up to down. I find it strange that such (large) changes can happen almost
instantaneously. But that's what quantum mechanics predicts..." This example again illustrates the need for helping students build
schemas consistent with quantum mechanics.

We performed a study in which we asked students: ``If the period $T=2 \pi/\omega$ of a Quantum Harmonic Oscillator (QHO) is known, how would you verify it 
experimentally?" 
A common response was that one must figure out some way to perform two successive measurements of the position of a particle 
separated by a time $T$ to make sure they are the same. 
The problem is that according to the postulate of quantum mechanics one cannot measure the position twice for the same particle and
hope to get the desired information because the first measurement will collapse the state. Further probing during interviews showed that students
were transfering the classical notion of period using the logic that the position of the QHO is deterministic. 
Interviews clarified that students were not referring to the ``probability"
of position measurement repeating itself after a time $T$ or to the coherent states of a QHO.
Few students provided correct response along the line that one can perform measurements of position
on ensembles of identically prepared systems, e.g., measure position on half of the systems at time $t=0$ and half after time $T$. 
The expectation values of position obtained in the two cases should be the same. 
Another correct response involving an experiment to obtain the energy separation between
the adjacent levels to find $\omega$ and using it to determine $T$ was rare.

Written responses and individual interviews also show that students have difficulty distinguishing between classical and quantum bound states.
Classically a particle is bound if the energy of the particle is such that there are two turning points. Quantum bound state is different due
to the quantum mechanical tunneling and the energy of the particle in a bound state must be less than the potential at both $-\infty$ and $+\infty$.
A majority of the students transfered the classical notion and looked for classical turning points to determine quantum bound states.

\section{Inappropriate Transfer from Modern Physics}

Experts understand the limits of the models they employ and use more refined models when simpler ones fail. Modern physics courses
often emphasize semi-classical models without helping students understand their limitations. For example, WKB approximation is used
as a back-of-the-envelope calculation for 1-D Time-Independent Schroedinger Equation (TISE). Within this approximation, instead of
an extended wave function with varying probability density in regions with different potentials, one often uses language such as ``The particle
spends more ``time" in this region since its total energy is fixed but if the potential energy is higher then the kinetic energy is lower".
Students who do not understand the limits of these models take the word ``time" literally and believe that there is a non-trivial 
time-dependence even in a stationary state. These notions from semi-classical model of what ``time-development" means makes learning
time-development in QM even more challenging. Further misconceptions that develop from this include ideas such as ``There must be work done
on a particle when it tunnels through a barrier because the particle was initially on one side of the barrier and later on the other
side so it must dig a hole to get through"~\cite{morgan}. Modern physics texts often use a similar ``imprecise" language when describing an electron in the stationary 
state of a hydrogen atom. They write that the electron spends more ``time" in one place and less ``time" in another instead of explanation involving
the probability of finding it at different distances from the nucleus. The notion that the particle is either here or there but spends different 
amount of time in different places is inconsistent with the indeterministic nature of position in quantum mechanics. It breeds many misconceptions
that become very difficult to rectify in QM. 

Our research also shows that students have trouble distinguishing between individual measurements and expectation values of physical observables. 
Modern physics courses
often avoid discussion of the probability of measuring individual outcomes and only discuss expectation values. Many students
conclude that for a physical observable, its expectation value and individual measurement are the same or that each individual measurement of
the observable will yield the expectation value. 
Individual interviews also show that many students believe that one cannot perform individual measurements in quantum
mechanics despite the fact that they have learned about double slit experiments with single photon or single electron. In such experiments, 
interference pattern builds up gradually as particles hit the screen one by one. It is clear from such experiments, that individual measurement
of position of the particle on the screen cannot be predicted ahead of time and individual measurement is very different from the expectation value of position.
Our research also shows that students often do not understand that expectation value refers to the average value of the observable
obtained after performing
a large number of measurements on identically prepared systems and have difficulty understanding the active nature of measurement in quantum mechanics.

Students in QM learn about incompatible observables (position-momentum, position-energy etc.) and the uncertainty relation between them.
After relevant instruction, we asked students about the measurement of the distance from the nucleus of an electron in the ground state of a 
hydrogen atom.
Many students claimed it will always yield the same value and some explicitly cited the Bohr model.

\section{Difficulty in Lateral Transfer}

Quantum postulates are very different from everyday experience and there is a ``quantum leap" in the mathematical sophistication required.
Therefore, it is difficult for students to de-contextualize concepts and apply them in contexts other than the ones in which they were acquired. 
For example, the superposition principle which is learned previously in many classical contexts
is also crucial for understanding the wave character of particles, in particular for grasping how the wave-packets are formed.
In the interviews, we found that even those QM students who can explain the superposition 
principle in classical contexts have difficulty in discerning its relevance to ``particle" waves. 
Many believe that for a given potential, the stationary states are the only wave functions possible. They have trouble understanding that 
any smooth wave function meeting the boundary condition is acceptable and wave packets are formed by the linear superposition of stationary states.
The Time-Dependent Schroedinger Equation (TDSE) is mentioned in passing in most modern physics texts and the idea of
linear superposition of wave function is usually ignored. In modern physics, the concept of wave packet is introduced in an ad hoc way when 
discussing the semi-classical models and wave-particle duality but almost never after the quantum mechanical model is introduced (except in the 
context of tunneling through barriers where little justification or reasoning is provided for why wave-packets are valid states of the system).
Unfortunately, TDSE is severely deemphasized even in QM courses. 

We posed to QM students
a question about whether a particle in a 1-D infinite square well potential of width ``a" can be in a state $sin^3(n\pi x/a)$ at a given 
time.
Students had learned in the lectures and homeworks that the particle can be in the states $A sin(\pi x/a)+B sin(2\pi x/a)$ or $Ax(a-x)$
at a given time since they represent linear superposition of stationary states and satisfy the relevant boundary conditions. Many students claimed that $A sin(n \pi x/a)$ were
the only possible solutions. When we explicitly posed a question asking whether they agree with a student who claims that all allowed wave functions satisfy $H\Psi=E\Psi$, roughly one third 
said ``yes". Some explicitly claimed that this is true ``by definition". Even when the question was posed explicitly in the context of hydrogen atom about whether the stationary states were the only possible 
wave functions, the response was similar. Transferring superposition ideas from one type of potential to another type (e.g., from infinite square
well to QHO) was also challenging for students. 

Time-development of wave function is an important concept for understanding among other things,
magnetic resonance imaging, interaction of light with matter and quantum computing. In a linear superposition of stationary states, each term
evolves in time via its own phase factor, e.g., $A exp(-iE_1 t/\hbar)\Psi_1+B exp(-iE_2 t/\hbar)\Psi_2$. Students have difficulty in transferring
these concepts learned in one context to another because the instruction does not help them recognize the deep features involved in
quantum dynamics and the role of the system Hamiltonian in this process~\cite{singh}. Our research shows that the most common mistakes in the questions asking for
the wave function after time $t$, given the wave function at $t=0$, are responses such as $(A\Psi_1+B \Psi_2)exp(-iE t/\hbar)$ or 
$Ax(a-x)exp(-iE t/\hbar)$. These expressions are
not solutions of TDSE and cannot represent valid wave functions. During the interviews, when students who provided such responses were asked about the meaning of ``$E$", the common response was
that it is the energy of that ``state". This statement is incorrect because 
these are not states of definite energy.

In another study, we asked students to explain whether the probability density should depend on time if the wave function of a particle in
an infinite square well at time $t=0$ was $A \Psi_1+B \Psi_2$. Interestingly, even those who recognized that each stationary state evolves
according to different time-dependent phase factors, often said that the probability density should be time-independent. They claimed that
the time dependence will cancel out when one takes the absolute square of the wave function. Their explanations show that
they lost the interference effects because they were adding the probabilities of being in each stationary state ($\vert A \Psi_1 \vert^2 + \vert B \Psi_2 \vert^2$) 
instead of adding the probability amplitudes at time $t$ and then squaring. It is important to give students an opportunity to think through
the consequences of this analysis in the context of concrete experiments, e.g., a double slit experiment.

Students often over-generalize results related to
``position" to other observables. For example, quantum measurement hypothesis says that the measurement of an observable will collapse the state
into an eigenstate of the corresponding operator. Our research shows that many students believe that the measurements of all physical observable, e.g., energy,
will collapse the wave function into a position eigenstate (a delta function in position). Students also struggle with qualitative ideas behind the
Fourier transform. We asked students the following question: ``Given that at time $t=0$, the wave function $\Psi(x)$ of a particle
incident on a potential barrier is a localized wave packet approximated by a delta function, what if anything can you say about the momentum
space wave function $\tilde \Psi(k)$? 
Draw a rough sketch with your explanation." The common responses claimed that $\tilde \Psi(k)$ should also be localized about the same location,
but perhaps be slightly broader or that one should just find the ``$k_0$" corresponding to the ``$x_0$" where the particle is localized
(some sketched square barriers in k-space right below the barrier in x space).

Our investigations also show that students do not automatically transfer the formalism of quantum mechanics learned in the 2-D Hilbert
space of spin 1/2 system to the infinite dimensional Hilbert spaces.
Even if students have
internalized the formalism for spin 1/2, connection with the infinite dimensional space is challenging for at least two reasons. First of all, 
the mathematical complexity increases manifold. Secondly, the infinite dimensional Hilbert space dealing with the orbital degrees of freedom
causes greater interference with students' cherished classical ideas. Quantum mechanical spin of particles is abstract and students do not
have much ``feel" for it. Therefore, concepts such as ``the spin of the particle does not in general have a definite value" or ``different 
components of spin cannot be measured simultaneously" do not produce as much conflict with the existing schemas. On the
other hand, it is much more difficult to internalize that position, momentum or energy of the particle do not in general have definite values or
that they cannot be measured simultaneously. The existing classical schemas related to 
these observables interferes with the counter-intuitive quantum mechanical tenets.

\section{Summary}

In order to help students transfer concepts from the previous courses to QM or within QM from one context to another appropriately, 
it is important to give them opportunity to build schemas consistent with the postulates of quantum mechanics and help establish lateral connections 
between useful concepts learned previously
and the new knowledge.
Our research also shows the need for better alignment between instruction in modern physics and QM. In particular, the manner and extent to which
semi-classical models are covered in modern physics should be re-evaluated. 

One encouraging finding from the perspective of developing effective instructional strategies is that despite the abstractness of quantum mechanics,
the types of difficulties students have are universal. Students over-generalize classical and semi-classical ideas in ways that can
be classified in a few categories. We are developing and assessing Quantum Interactive Learning Tutorials (QuILTs) that take into account student 
difficulties and incorporate paper-pencil tasks and computer simulations. 
The research-based tutorials that provide scaffolding.
They are designed to help students build robust schemas consistent with
quantum mechanics principles while helping them discern how their prior knowledge resources can be utilized in appropriate situations.
They help students organize their knowledge hierarchically 
and represent it at a more abstract level in memory to facilitate de-contextualization, categorization and recognition based upon deep quantum
mechanics laws.

\begin{theacknowledgments}
We are grateful to the NSF for award PHY-0244708.
\end{theacknowledgments}

\bibliographystyle{aipproc}   
{}

\IfFileExists{\jobname.bbl}{}
 {\typeout{}
  \typeout{******************************************}
  \typeout{** Please run "bibtex \jobname" to obtain}
  \typeout{** the bibliography and then re-run LaTeX}
  \typeout{** twice to fix the references!}
  \typeout{******************************************}
  \typeout{}
 }

\end{document}